\documentclass[epj]{svjour}

\usepackage{graphicx}
\usepackage{fancyhdr}
\def\makeheadbox{{
 }}
\def\mytitle{My title} 
\def\myauthors{My name}  
\def\mytype{My type of session}
\def\mysession{My session}

\def\mytitle{SUSY@LHC.CERN.CH} 
\def\myauthors{Maria}    
\def\mytype{Plenary}
\def\mysession{\myauthors}

\def\met{\mbox{$E_{\rm{T}}^{\rm{miss}}$}}
\newcommand{\meff}     {$M_{\mathrm{eff}}$}
\newcommand{\mt}       {$M_{T}$}
\newcommand{\st}       {$S_{T}$}
\begin{document}
\title{SUSY@LHC.CERN.CH}
\subtitle{Search, Navigation and Orientation}
\author{Maria Spiropulu\inst{1}
\thanks{\emph{Email:} smaria@cern.ch}%
}                     
%
%
\institute{Physics Department, CERN, CH 1211 Geneva 23}
%
\date{\today}
\abstract{
%
I discuss the program of work towards discoveries
at the LHC, and I include seeds for orientation and navigation
in the parameter space given the foreseen multitude of excesses at startup.
} 
\maketitle
%

\section{Introduction}
\label{intro}
The Large Hadron Collider will produce 14 TeV proton-proton collisions 
in probably less than a year from when this proceeding is published. 
ATLAS and CMS,  are focusing this year
on the final commissioning of the experiments and what I call ``engineering
the discovery plan''. The strategies for the careful understanding and use 
of the Standard Model data at 14 TeV constitutes a large part of the 
readiness for discovery at the LHC.

While a number of modern theoretical  frameworks have
emerged in the past decade, most all dual to the previous canonical 
beyond-the-standard physics ideas and models, 
supersymmetry appears to  still have no rivals as the 
top and favorite theory that embraces and
enhances the Standard Model at the TeV scale. 
In fact a lot, if not most, of the models implied above 
end up looking eventually like SUSY at the TeV scale
 (UEDs, little-Higgs with T-parity etc). The rest postpone the
introduction of TeV new physics to multi-TeV new physics. 

I will not indulge in the theoretical reasons of why when we try to extend 
the Standard Model at short distances, as short as the Planck length,
we need the introduction of new theories.  I would instead 
like to remind ourselves that with the  Standard Model we close
a more than two thousand years cycle of theoretical and experimental exploration into the nature of matter and its interactions. The Standard Model
is extremely successful and precise, to one part in a billion in many cases. 
Together with the general theory of relativity it is fair to say that
we have a correct theory of the known fundamental constituents of matter
and their interactions down to length scales of $10^{-18}$ cm. This
by no means implies that we understand the physics mechanisms by which the 
Standard Model and its contents emerge the way we observe them in the experiments.

Supersymmetry (\cite{Wess:1974tw},\cite{Fayet:1976cr},\cite{Barbieri:1987xf},\cite{Hall:1983iz}) was initially constructed to help introduce fermions in string theory (\cite{Ramond:1971gb}); string theory itself was built to
describe the quark interactions (e.g. gluonic flux tubes;   the jet ``strings'' in the printout of a PYTHIA event is not a coincidence see e.g. \cite{Andersson:1983ia},\cite{Sjostrand:1982fn},\cite{Gross:1973id},\cite{Politzer:1973fx},\cite{Field:1977fa}).
Forty years of experimental results from accelerators, astrophysical 
and cosmological observations and  progress in theory
are pointing to  LHC's likelihood of discovering new physics. 

Two are the major experimental observations  that
in concert with the theoretical considerations can be used as corroborative 
evidence for physics mechanisms that broaden the Standard Model:

\begin{enumerate}
\item the observed dark matter in the universe
\item the observed masses of the $W$ and $Z$ vector bosons
\end{enumerate}

The expectation is then that the LHC will discover a new sector of 
particles/fields associated with electroweak symmetry breaking and 
dark matter. Supersymmetry outputs both and is the best template of discovery
physics. Note that indeed we don't know {\it a-priori} what the discoveries
will be. Preparing for the discoveries ahead of time given the best 
templates does not guarantee nor does it imply that these exact template(s) 
is what we (expect) will be found,  
nor that the preparation strategies are sufficient and exact to assist the
discoveries, come data time. It only implies that we investigate in detail
all we (think we) know, and think well on all we know we don't know.

\section{The  program of work}
\label{sec:1}
\subsection{Status of the experiments}
Both the ATLAS (A large ToroidaL ApparatuS) and CMS (Compact Muon Solenoid)
are in stage of commissioning. Already both experiments are collecting 
astrophysics data and finalizing the analysis of beam tests data of most-all 
detector elements.The details of their everyday progress, as well as the 
status of the accelerator can be found at the corresponding CERN sites. 
The expected performance of the experiments will be published in early 2008.
According to the published schedule of the lab (also see the LHC plenary talk in this meeting by Lyn Evans \cite{lyn}) we expect 14 TeV collisions before the end of 2008. While in what follows I focus on the searches for supersymmetry at ATLAS and CMS I must point out that the discovery of supersymmetry only emphasizes the many flavor mysteries that can only be resolved in dedicated flavor experiments, many of which can only be performed at LHCb \cite{Lenzi:2007nq},\cite{Blouw:2007px}.

\subsection{Outline of work towards early discoveries}
The preparatory/readiness work on early sypersymmetry targeting  discoveries  
at  ATLAS can be summarized as follows:
\begin{itemize}

\item Data-driven Estimation of $Z/W$ background to SUSY
\item Data-driven Estimation of top background to SUSY
\item Data-driven Estimation of QCD background to SUSY
\item Estimation of Heavy Flavor backgrounds and associated systematic
\item Searches and inclusive studies for SUSY events
\item Exclusive measurements for SUSY events
\item Gaugino direct production
\item Studies for gauge-mediated SUSY

\end{itemize}
Similarly the corresponding  CMS program of work is organized as follows:
\begin{itemize}
\item  Leptonic searches (MSSM template)
     \begin{itemize}
         \item \small Search for SUSY in $\ge$1 lepton+$E_{\rm T}^{\rm{miss}}$ + jets at 14 TeV  in the electron and muon channels ($\mathcal O$(100 $pb^{-1}$)).  
         \item \small in dilepton pairs+ $E_{\rm T}^{\rm{miss}}$+jets at 14 TeV  in the electron and muon channels ($\mathcal O$(100 $pb^{-1}$)).
         \item \small Search for SUSY in trileptons + jets at 14 TeV.  (1 $fb^{-1}$).
      \end{itemize}
\item Hadronic searches (MSSM template)  
       \begin{itemize}
          \item \small Search for SUSY in   0 lepton + $E_{\rm T}^{\rm{miss}}$ + jets at 14 TeV  ($\mathcal O(100)$ pb$^{-1}$).
           \item \small in $b\bar{b}$ + $E_{\rm T}^{\rm{miss}}$ + jets at 14 TeV ($\mathcal O(100)$ pb$^{-1}$). 
       \end{itemize}
\item  Heavy Stable Charged Particles and photonic searches (GMSB template) 
        \begin{itemize}
        \item \small Search and reconstruction of heavy stable charged particles at 14 TeV using TOF and dE/dx (500 pb$^{-1}$, model dependent). 
        \item \small Search for GMSB using prompt photons at 14 TeV (500 pb$^{-1}$).
        \end{itemize} 
\end{itemize}

To orient ourselves in the vast theoretical parameter space, we expect an iterative process of investigative work once the data show excesses that can be
briefly outlined as follows (\cite{SUSY06}):

\begin{itemize}
\item choose well-defined inclusive signatures
\item extract some constraints on masses, couplings, spin from decay 
kinematics and rates 
\item try to match emerging pattern to tentative template models
\item having adjusted template models to measurements, try to find additional signatures to  discriminate different options 
\end{itemize}

This program of work calls for  ``realistic'' analyses that prepares the experiments as thoroughly as possible for the  real data analyses.
It implies identifying  and implementing   the crucial groundwork in terms of detector  understanding, physics object requirements, 
trigger understanding and requirements, dataset definitions, and 
potential systematic uncertainties especially at startup. Of particular gravity
is the development of  methods for extracting backgrounds
and particle identification efficiencies from data wherever possible, 
and the definition of  trigger paths  (for a summary of the status of trigger at the ATLAS and CMS  experiments see \cite{DeSanto:2007ta}) and datasets needed for these measurements. 

Both experiments are also carrying out detailed studies aiming at  non-supersymmetric exotic model signatures and searches.

\section{Discovery signatures}

To comply with the measured proton lifetime ${\cal O}(10^{33}{\mathrm yrs})$, a what seems to be {\it ad-hoc} symmetry is introduced to generic minimal supersymmetric models: $R$-parity, $R=(-1)^{3(B-L) + 2 s}$, where for each particle
$s$ is the spin, and $B$ and $L$ are the respective baryon and lepton
assignments. The consequence of R-parity conservation is a stable lightest supersymmetric particle (LSP) that in most of the models is weakly interacting and provides a fair candidate for a component of the observed dark matter in the universe. Due to the pair of LSPs a characteristic ensemble of signatures contains large missing energy along with high number of jets and leptons. I will highlight some important aspects of analyses related to this ``vanilla'' type of SUSY searches in what follows. Signatures and searches associated with GMSB or split-SUSY frameworks are reviewed in this meeting and summarized in \cite{Zalewski:2007up} \cite{Bressler:2007gk}.

\subsection{All-hadronic final states with large missing energy}

\begin{figure}[htb]
\begin{center}
    \includegraphics[width=.5\textwidth]{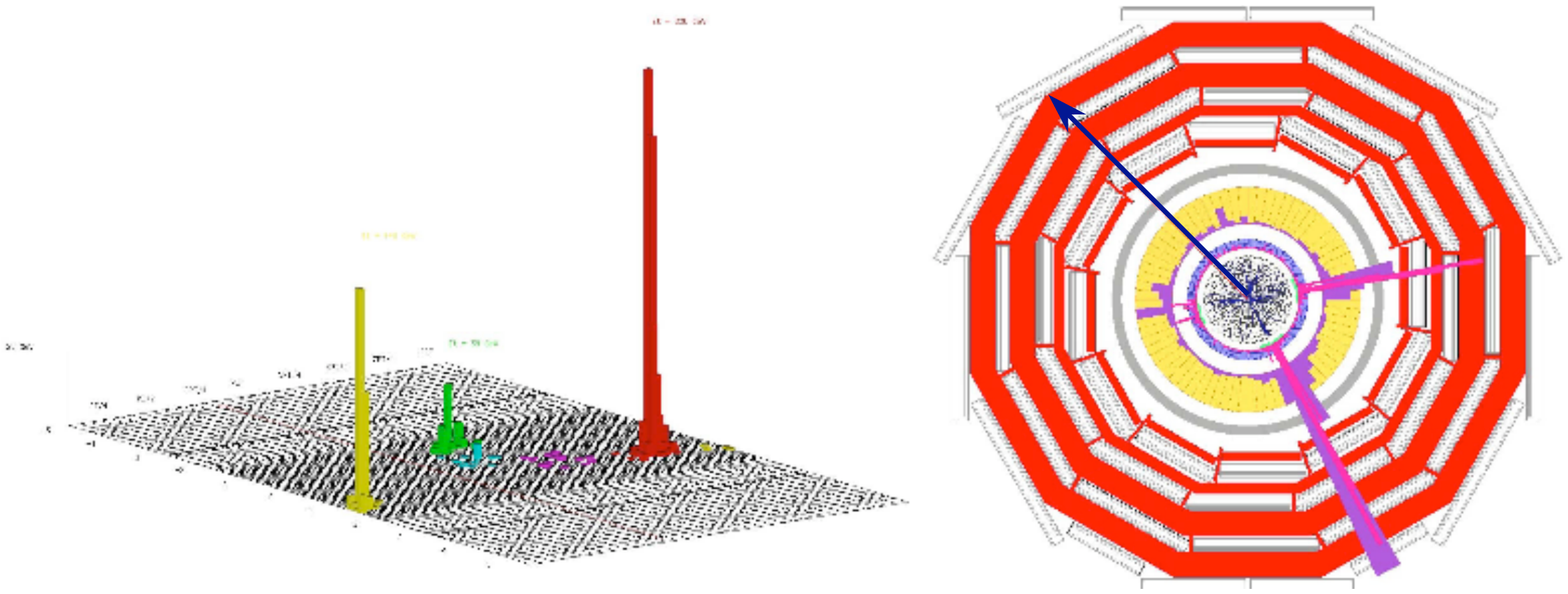}
\end{center}
    \caption{Event display of SUSY candidate event that survives the requirements  of the CMS multijet+missing energy
analysis. The three highest $E_T$ jets are 330, 140 and 60 
GeV while the missing transverse
energy is 360 GeV. (left) Lego $\eta-\phi$ calorimeter display, the 
three leading jets
are color coded red-yellow-green, while the missing energy $\phi$ is 
indicated with the
red line  (right) transverse $x-y$ view, shows relative depositions 
of the jets in the calorimeter systems
as well as the reconstructed tracks and the missing energy vector 
direction.}
    \label{fig:candidate}
\end{figure}

The canonical search and discovery of gluinos and squarks  
is using the large missing transverse energy plus multijet signature. 
The large missing energy  originates from the two LSPs in the final 
states of the squark and gluino decays. The three or more  hadronic jets  
result from the hadronic decays of the $\tilde{q}$ and/or $\tilde{g}$.
Such an event display at the CMS detector is shown in Figure ~\ref{fig:candidate}. 
The search proceeds in a dataset triggered by missing energy and jets,
a legentarily notorious dataset in hadron colliders plagued by all types of 
instrumental and spurious backgrounds. Clean-up methods that invoke the event electromagnetic fraction and event charged fraction as first designed at the Tevatron \cite{mine} are also employed here - the final demonstration of their
effectiveness is under study with the detailed simulation of beam halo and cosmic events for example, where the techniques proved to be particularly 
efficient at the Tevatron. 

Due the very high QCD production cross section the SM background to
a large missing transverse energy plus jets data-sample is dominated
by QCD production. The observed missing transverse energy in QCD jet
production is largely a result of jet mismeasurements and detector
resolution. Methods to eliminate QCD events based on
angular correlations between the jets and the missing energy are employed
as summarized in \cite{Tytgat:2007gj} and the effects of the jet resolution 
on the tails of the missing energy distribution  at \cite{Yamamoto:2007it}
and \cite{Yetkin:2007zz}. 

\subsubsection{``Standard Candle'' Calibration}

The so-called ``standard candle calibration'' methods are pivotal in extracting the Standard Model background normalization and shapes from the data, in particular with the early data \footnote{Because of their extreme brightness, type Ia supernovae have become part of the cosmological tool kit as "standard candles" used to measure distances to galaxies; we borrow the nomenclature when using clean standard model signals to normalize background predictions to new physics.}. They have also been shown to provide robust predictions  in searches at the Tevatron \cite{mine}. In what follows I discuss in detail a major standard model candle, the $Z^{0}$ boson. 

Events with large missing transverse energy and
$\ge$3  jets in the final state are expected from
$Z(\rightarrow \nu\bar{\nu})+\ge$ 3 jets and $W(\rightarrow
\tau\nu)+\ge$ 2 jets (the third jet originating from the hadronic
$\tau$ decay) processes. Additional residual contribution
is expected also from $W(\rightarrow
\mu\nu), ~e\nu +\ge$3 jets.   Both ATLAS and CMS  are designing  a comprehensive
normalization program that relies on the $Z$  + multijet data 
(ATLAS also using the $W$+jets data) to 
accurately estimate the $W$ and $Z$+multijet background 
contribution in a large $E_T^{miss}$ plus multijet search.

The aim is to normalize the Monte Carlo predictions for events
with $\ge3$ jets and $Z$ boson $P_{T}>$ 200 GeV  to the observed
$Z (\rightarrow \mu\mu)+$ 2 jets data sample ( where $Z$ boson $P_{T}>$ 200 GeV  ) via the measured $R = \frac{dN_{events}}{dN_{jets}}$  ratio.

\begin{figure}[htb]
\begin{center}
    \includegraphics[width=.45\textwidth]{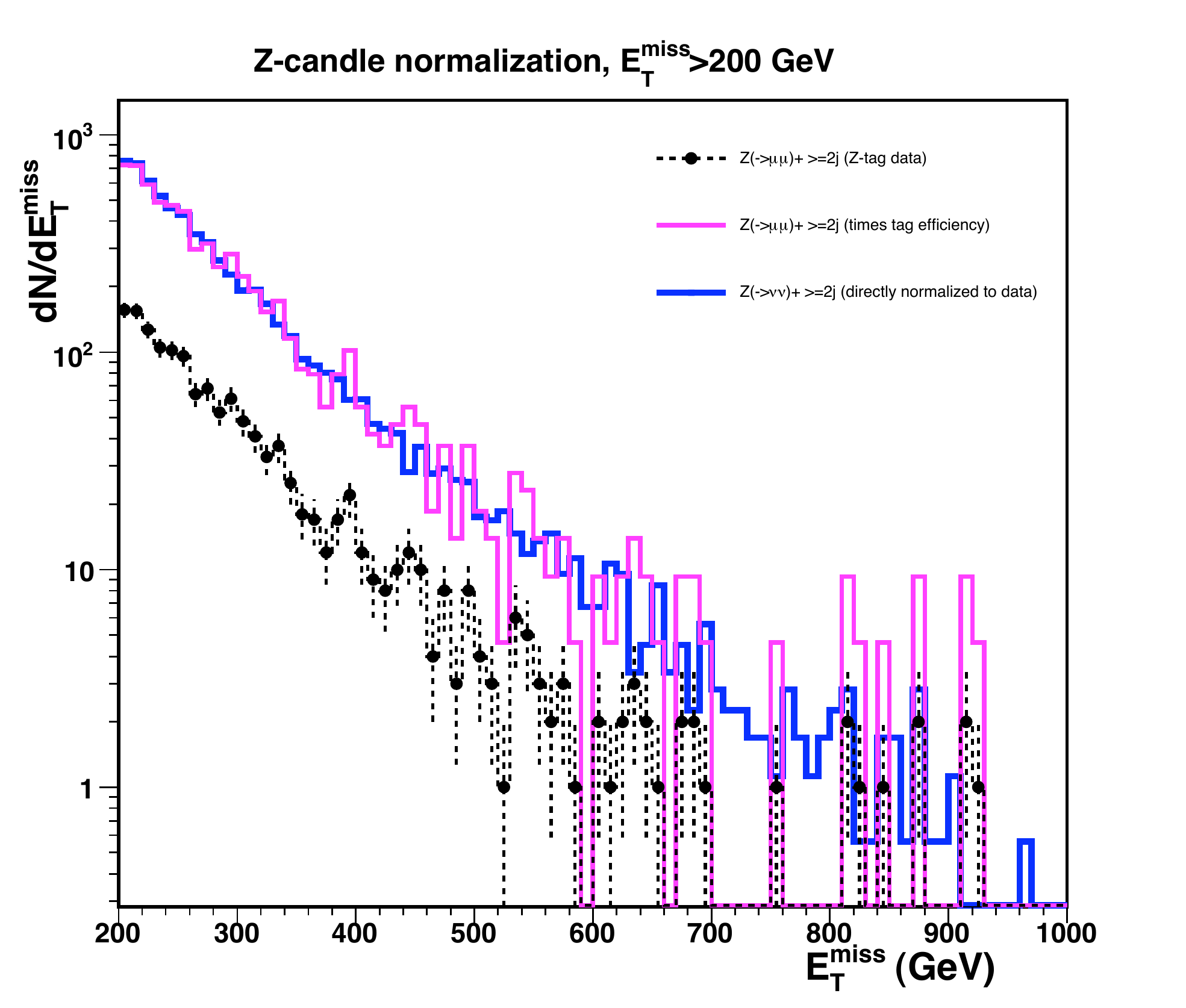}
\end{center}
    \caption{\met~ in  $Z\rightarrow\mu\mu$ + $\ge$ 2 jets candle sample and normalized   \met~  in  $Z\rightarrow\nu\bar{\nu}$ + $\ge$ 2 jets sample.}
    \label{fig:metmunu}
\end{figure}

\begin{figure}[htb]
\begin{center}
    \includegraphics[width=.45\textwidth]{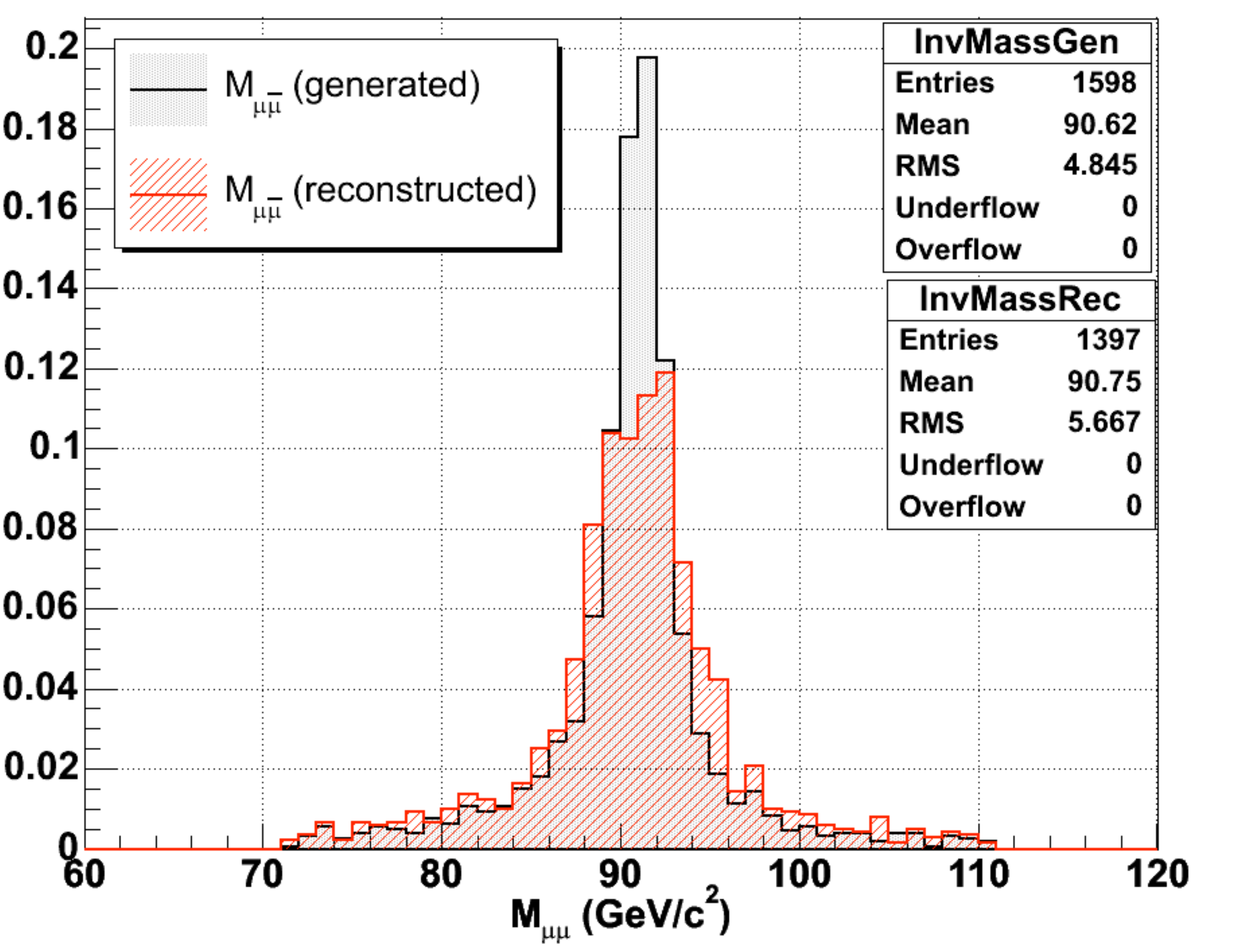}
\end{center}
    \caption{Reconstructed  and generator level $Z$ dimuon invariant mass
for $Z\rightarrow\mu{\mu}$ + $\ge$ 2 jets and  $\met>$ 200 GeV. }
    \label{fig:Zmass}
\end{figure}

As an example the $Z\rightarrow \mu\mu$ +$\ge$ 2 jets with $Z_{PT}>200$ GeV
is used as the  ``candle'' data sample. 
The selected candle sample dimuon invariant mass
is shown in Figure \ref{fig:Zmass} overlaid with the one using the Monte Carlo truth. Both the muon and electron decays of the $Z$ will be used as the standardizable candle, but for the purposes of demonstrating the method,  the $Z$ muon decays are chosen.  Since the rudimentary calorimetric 
missing transverse energy is used (as is likely to be the case at
the start-up of the experiment),  the shape of the
$E_T^{miss}$ distribution of the measured
 the $Z\rightarrow \mu\mu$ +$\ge$ 2 jet events  will be
very close to the shape of the 
invisible $Z\rightarrow \nu\nu$ +$\ge$ 2 jet events
as shown in Figure \ref{fig:metmunu}. 


The ratio 
$\rho\equiv\frac{\sigma(pp \rightarrow W(\rightarrow \mu(e)\nu)+jets)}{
\sigma(pp\rightarrow Z(\rightarrow \mu^{+}\mu^{-})(e^{+}e^{-})+jets)}$ will be used
to normalize the $W$+jets Monte Carlo predictions.
Assuming lepton universality,
the predictions for the number of events with $\ge 2$-- and $\ge
3$--jets from $W$ and $Z$ production and decays to all flavors will be
normalized to the  $Z(\rightarrow \mu^{+}\mu^{-})+\ge 2$ jets data.  By
normalizing the MC predictions to data systematic
effects in particular at the early data taking stages can be ameliorated.

While the $Z$ boson provides a very clean normalization candle  both ATLAS and CMS are
designing  the strategy for the extraction of the top background at start-up also using the data and not relying on the Monte Carlo predictions. The top (see e.g. in \cite{Tytgat:2007gj}) as well as the $W$ provide less clean standard candles (due to ambiguities in their mass reconstruction) but at the LHC their production rate is very high and their role in the discovery plan will be crucial. In all cases the tails of the Standard Model processes such as $W$, $Z$, and top QCD associated production, will be enriched with SUSY signal events and the full standard candle program needs to 
demonstrate robustness against normalizing away the probable signal.  The caveats and alerts on QCD associated production at the LHC and the use of the predictions are discussed extensively in the plenary talk and corresponding work  of Michelangelo Mangano \cite{MLM}. 

\subsubsection{Analysis paths for all-hadronic searches}

An ATLAS all-hadronic analysis path proceeds as follows:
\begin{itemize}
  \item $N_{\mathrm{jet}} \ge 4$,
  \item $p_{T}^\mathrm{J1} > 100~\mathrm{GeV/c} \ \& \ p_{T}^\mathrm{J4} > 50~\mathrm{GeV/c}$,
  \item $S_{T} > 0.2$,
  \item $E_{T}^{\mathrm{miss}} > 100~\mathrm{GeV} \ \& \ E_{T}^{\mathrm{miss}} > 0.2 \times M_{\mathrm{eff}}$,
\end{itemize}
where $N_{\mathrm{jet}}$, $p_{T}^\mathrm{J1(4)}$, \st~ and \meff~ are the number of jets , the transverse momentum of first (fourth) leading jet, the transverse sphericity and the effective mass, respectively.
The effective mass is defined as $ M_{\mathrm{eff}} = \sum_{i=0}^{i\le4} p_{T}^{i} + E_{T}^{\mathrm{miss}}$, where $p_{T}^{i}$ is the transverse momentum of $i$-th leading jet. The analysis path that includes leptons in the final state 
is similar with the additional selection of events requiring one isolated lepton with $p_{T}$ larger than $20~\mathrm{GeV}$ and the transverse mass \mt$>100~\mathrm{GeV}$.

A selection path for the all-hadronic CMS analysis is shown in Table 
~\ref{tab:path} with a remark indicating the reason  and aim of each 
selection step.
Notice that although the analysis is inclusive we introduce a number of steps 
targeting the cleanup of the dataset. These steps (e.g Event Electromagnetic Fraction (EEMF), Event Charged Fraction (ECHF)) are more than 90\%
efficient in the Monte Carlo studies both for the signal and the backgrounds 
but the are expected to eliminate instrumental spurious backgrounds in the real data. To reduce the large Standard Model background contribution mainly from $W(\rightarrow \ell\nu)+jets$, $Z(\rightarrow \ell\ell)+jets$ and $t\bar{t}$
production and decays an {\it indirect lepton veto} (ILV)
scheme is designed that uses the tracker and the calorimeter. 
The aim of the ILV is twofold: a) to retain large signal efficiency 
b) to achieve large rejection of the $W,Z,t\bar{t}$ backgrounds as shown 
in table ~\ref{tab:EvSel}. The final signal and background yield for 1 fb$^{-1}$
is given in table \ref{tab:res}.

\begin{table}[ht]
\begin{center}
\caption{The $\met$ + $\ge 3$ jets SUSY search analysis path.$H_{\mathrm{T}}=\sum_{i=2}^{4} p_{T}^{i} + E_{\mathrm{T}}^{\mathrm{miss}}$, for the event electromagnetic and charged fraction variables as well as the indirect lepton veto see \cite{Yetkin:2007zz}, \cite{Ball:2007zza}  {\label{tab:path}.}}
\begin{tiny}
\begin{tabular*}{0.45\textwidth}{ll}\hline\hline
Requirement & Remark \\ \hline \hline
Level 1 & L1 trigger efficiency \\
 & parameterization \\
HLT, $E_{T}^{miss}>$ 200 GeV & trigger/signal signature \\
primary vertex (PV) $\ge 1$ &  primary cleanup\\
$EEMF \geq$ 0.175, $ECHF\geq$ 0.1 & primary cleanup \\  \hline \hline
$N_{j}\geq$ 3,$|\eta_{d}^{1j}| < 1.7$  & signal signature \\  \hline \hline
$\delta\phi_{min}(\met -jet)\ge 0.3$ rad, &\\
$R1,R2>0.5$ rad, & \\
$\delta\phi( \met - j(2) )  > 20^{\circ}$ &  QCD rejection \\   \hline \hline
$Iso^{ltrk}=0$, & ILV (I)  \\
$EMF(j1),EMF(j2)<0.9$  & ILV (II), \\
 & $W/Z/t\bar{t}$ rejection \\ \hline\hline
$E_{T,j(1)} >$ 180 GeV,$E_{T,j(2)} >$ 110 GeV,  &  \\
$H_{T} > 500$ GeV & S/B optimization \\\hline \hline
\multicolumn{2}{c} {SUSY LM1 signal efficiency 13\%} \\ \hline\hline
\end{tabular*}
\end{tiny}
\end{center}
\end{table}

\begin{figure}[htb]
\begin{center}
    \includegraphics[width=.24\textwidth]{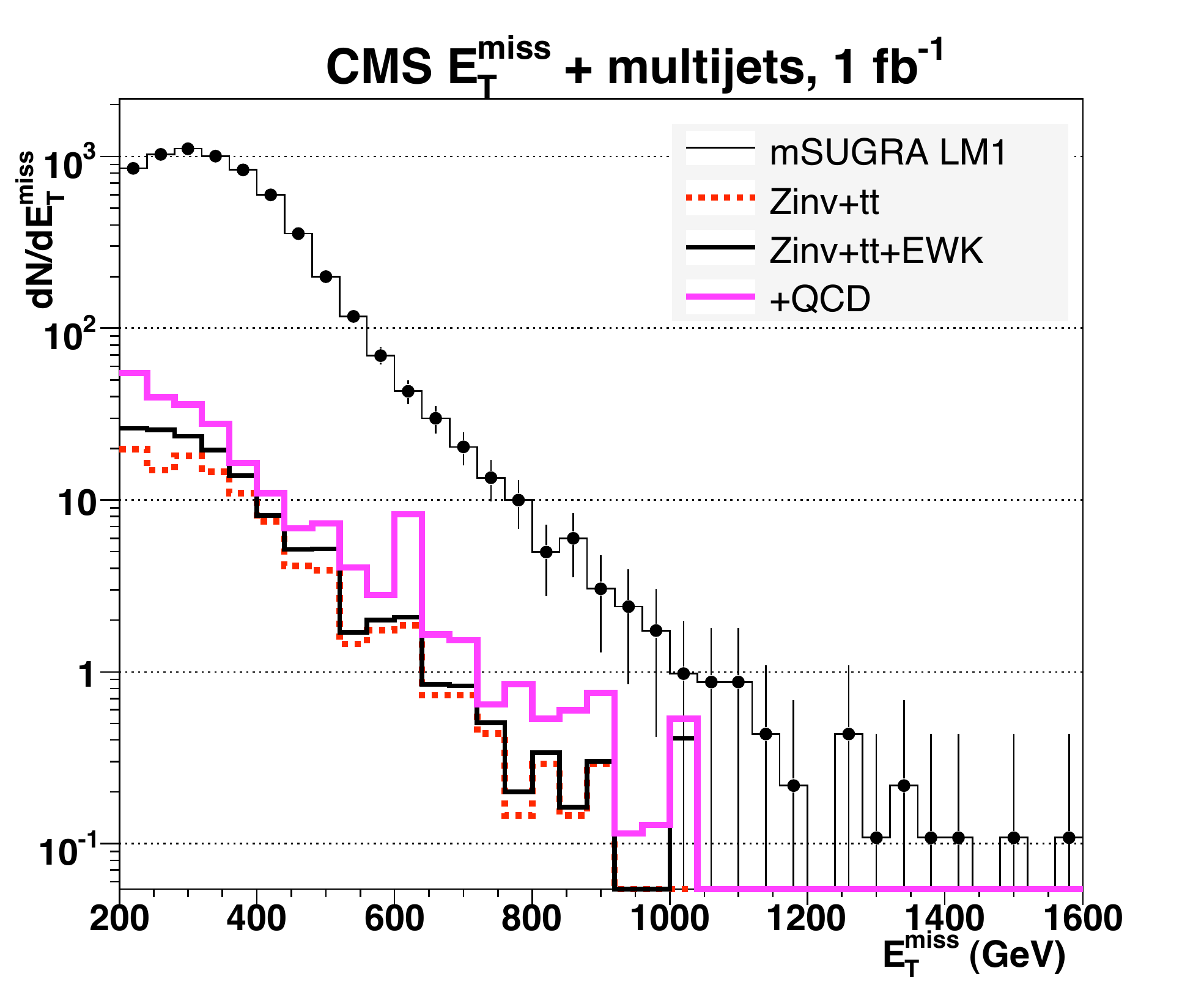}
    \includegraphics[width=.24\textwidth]{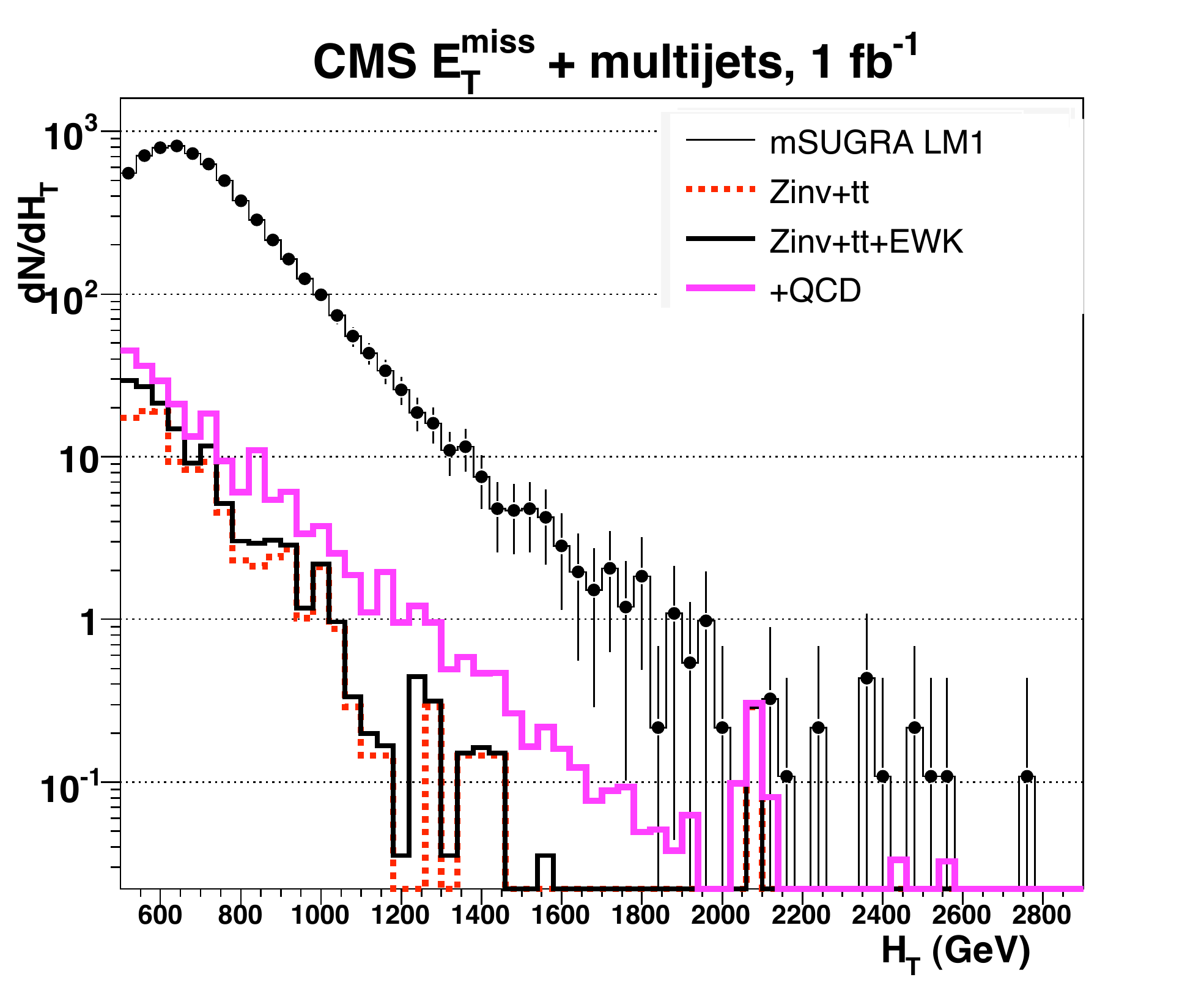}
    \includegraphics[width=.24\textwidth,height=0.21\textwidth]{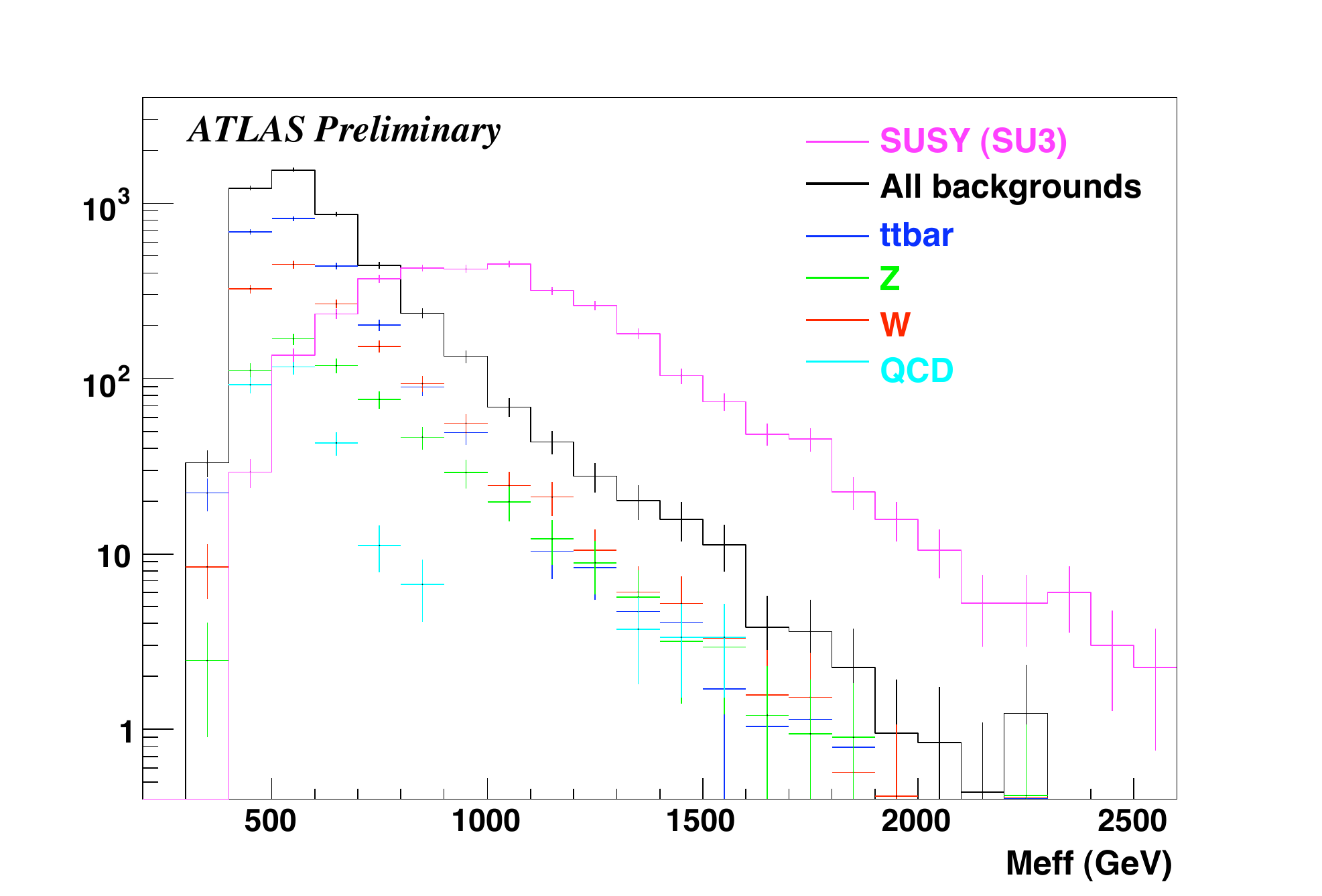}
    \includegraphics[width=.24\textwidth,height=0.21\textwidth]{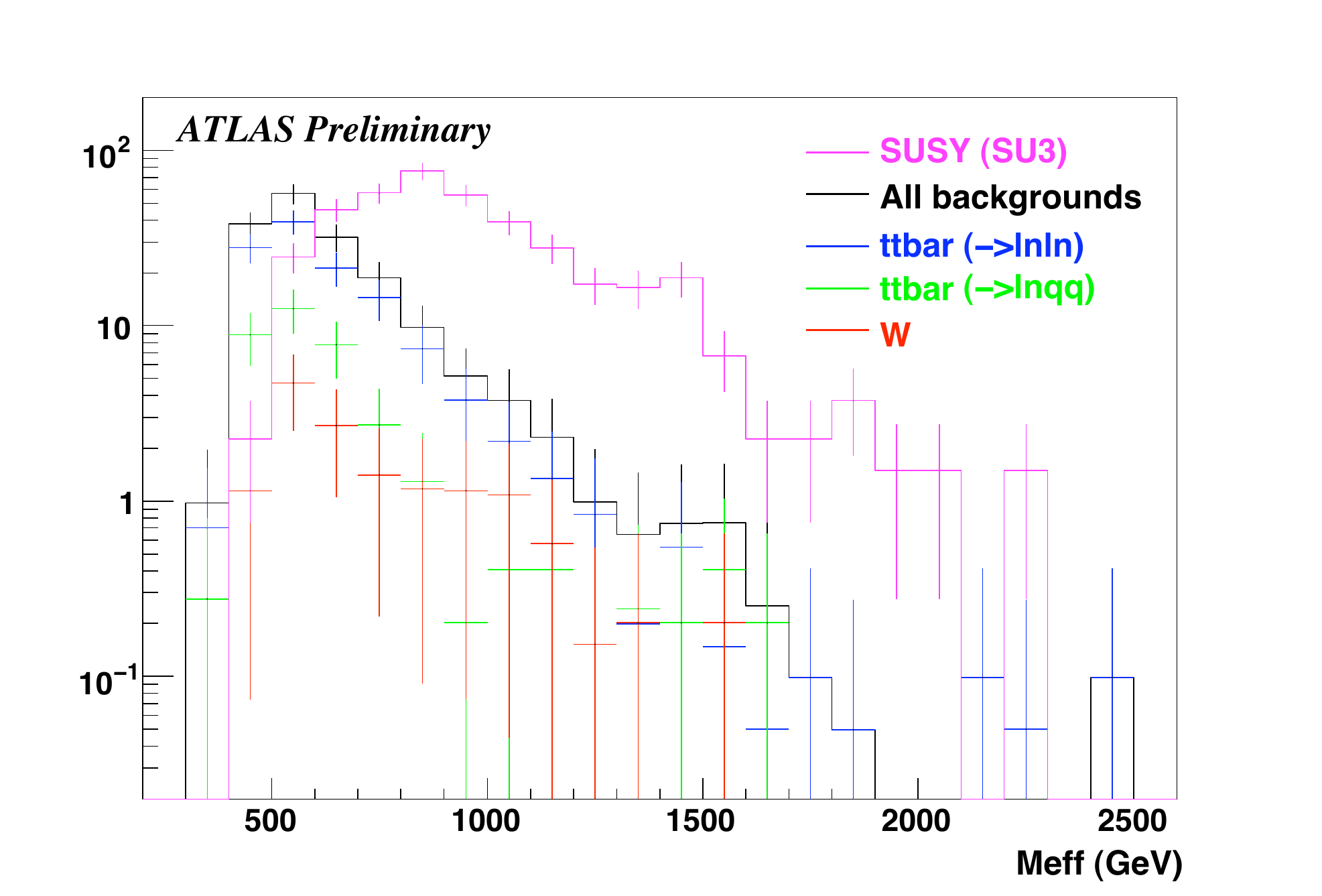}
\end{center}
    \caption{ (top) \met~ and $H_{\mathrm{T}}$ distributions in the all-hadronic CMS analysis. (bottom) The \meff~ distributions for no-lepton (left) and single lepton (right ) signatures in ATLAS.
All for integrated luminosity of $1\mathrm{fb}^{-1}$.}
    \label{fig:met-search}
\end{figure}

In Figure \ref{fig:met-search} the results are shown for the inclusive 
all-hadronic  \met+$\ge$3 jets search at CMS (top), the all hadronic
\met+$\ge$4 jets search at ATLAS (bottom left), and the \met+$\ge$4 jets + 1 
isolated lepton search at ATLAS (bottom right) for 1 fb$^{-1}$.  
 
\begin{table}[htb]
\begin{center}
\caption{Cumulative selection efficiency after each   requirement
in the $\met$+multijets analysis path for major Standard Model backgrounds.
(EWK refers to $W/Z$,$WW/ZZ/ZW$).}{\label{tab:EvSel}}
\begin{tiny}
\begin{tabular*}{0.45\textwidth}{lccccc}\hline\hline
Cut/Sample    & Signal & $t\bar{t}$   &  $Z(\to\nu\bar{\nu})+$ jets &  EWK + jets\\\hline
All (\%)           & 100 & 100     & 100                     & 100               \\ \hline \hline
Level-1       & 92 & 40                        & 99                         & 57 \\ \hline 
HLT           & 54 & 0.57                    & 54                         & 0.9 \\ \hline 
PV            & 53.8 & 0.56                  & 53                         & 0.9 \\ \hline 
$N_{j}\geq$3 & 39 & 0.36                    & 4                          & 0.1 \\ \hline 
$|\eta_{d}^{1st,j}\geq 1.7$ & 34 & 0.30       & 3                          & 0.07 \\  \hline \hline
EEMF $\geq$ 0.175 & 34         & 0.30        & 3                         & 0.07 \\ \hline 
ECHF $\geq$ 0.1   & 33.5        & 0.29       & 3                         & 0.06 \\ \hline \hline 
QCD angular & 26 & 0.17      & 2.5                       & 0.04\\ \hline \hline 
$Iso^{lead\;trk}=0$ & 23    & 0.09 &         2.3  & 0.02 \\ \hline 
$EMF(j1),$ & & & & &  \\
$EMF(j2) \geq 0.9$ & 22 & 0.086   & 2.2  & 0.02 \\ \hline\hline 
$P_{T,1} >$ 180 GeV, & & & & & \\
$P_{T,2} >$ 110 GeV  & 14  & 0.015  & 0.5        & 0.003 \\ \hline
$H_{T}> 500$ GeV & 13 & 0.01  & 0.4 & 0.002 \\ \hline\hline
\multicolumn{6}{c}{1/fb}\\\hline\hline
$H_{T}> 500$ GeV & 6319 & 53.9 & 48 & 33 \\ \hline\hline
\end{tabular*}
\end{tiny}
\end{center}
\end{table}

\begin{table}
\begin{center}
\caption{All-hadronic selected low mass SUSY and Standard Model
background events for 1 fb$^{-1}$ from CMS {\label{tab:res}}}
\begin{small}
\begin{tabular}{|l|c|}\hline\hline
 Signal (LM1) & 6319\\
 $t\bar{t}$/single $t$ & 56.5\\
$Z(\to\nu\bar{\nu})+$ jets & 48\\
 ($W/Z$,$WW/ZZ/ZW$) + jets & 33\\
QCD& 107 \\ \hline\hline
   \end{tabular}
\end{small}
\end{center}
\end{table}
Due to the QCD Monte Carlo limited statistics
to derive the QCD background component the analysis path
is followed without the topological QCD clean-up requirements
and ILV requirements.  The estimate is based
on factorizing the clean-up and ILV efficiency,
assuming them uncorrelated with the rest of the analysis requirements and using a parameterization of it as a function of the \met~ for the large $\met$
 tails.

\subsection{Leptonic signatures with large missing energy}

\begin{figure}[htb]
\begin{center}
    \includegraphics[width=.45\textwidth]{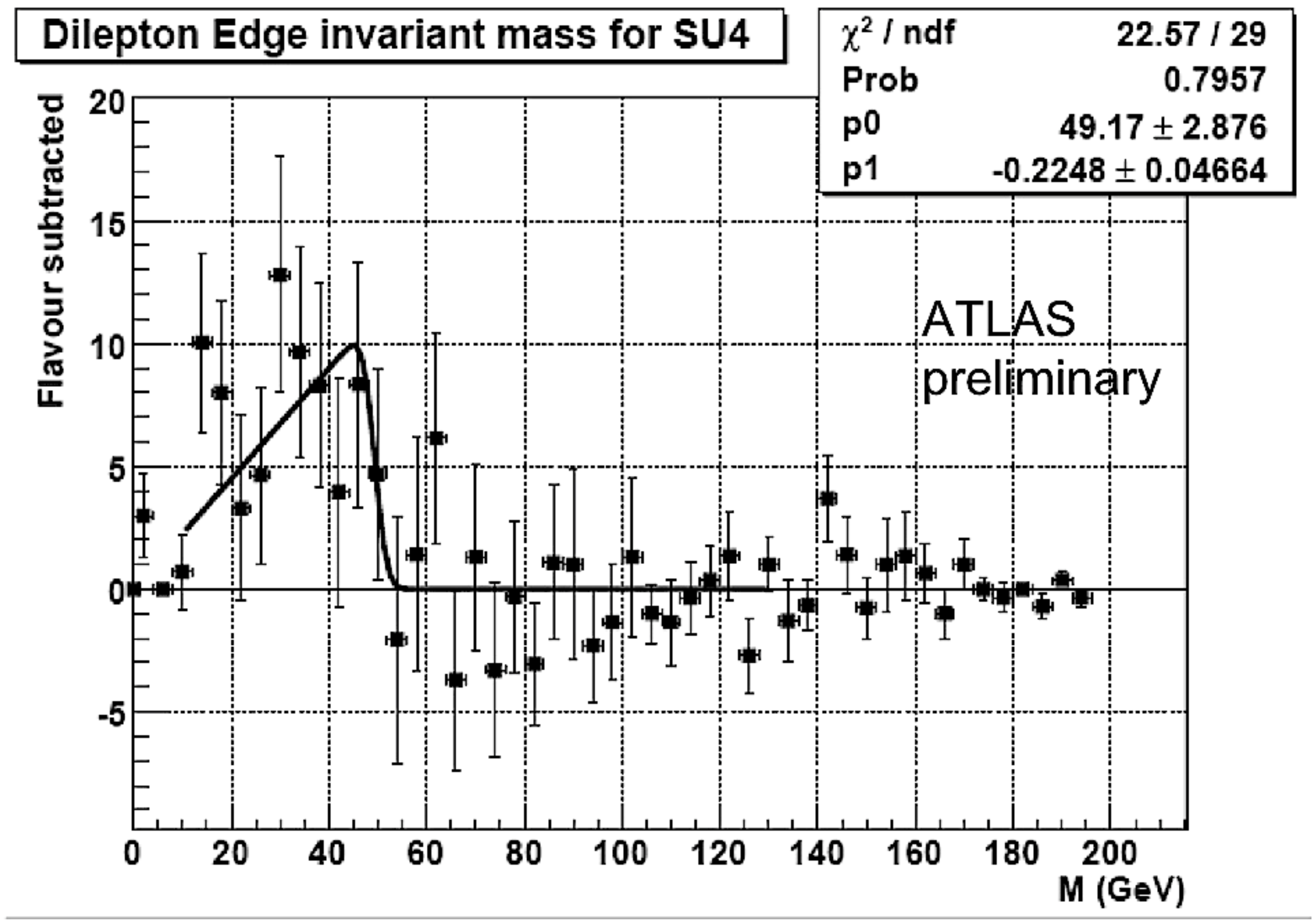}

\end{center}
    \caption{The dilepton invariant mass distribution for a full simulation sample of an ATLAS 
low mass benchmark SUSY point with an integrated luminosity of 350 pb$^{-1}$. A triangular function convoluted  with a Gaussian is used in the fit to estimate the edge position. Note that the signal significance is well over 5$\sigma$ significance with only 100 pb$^{-1}$ \cite{Ozturk:2007ap}.}
    \label{fig:su4}
\end{figure}

Signatures with leptons, jets and missing energy provide both discovery
and characterization channels for SUSY. 
Leptons are produced in the decays of charginos and neutralinos ;  their 
kinematic and topological characteristics as well as their mutliplicities 
including flavor and charge can point towards the production types and rates
(i.e.  mass hierarchies) of the squarks and gluinos and the composition of the LSP. Traditionally invariant  masses that involve dileptons and 
leptons+jets have been used at the LHC for  the mass reconstruction 
using large integrated luminosity. These studies are 
currently being worked for the early data and additional
 measurables are being introduced. 
The measurement and understanding of the trigger, 
lepton identification efficiencies and acceptance as 
well as fake rates are prerequisites for the lepton 
involving signatures to be rendered useful beyond the 
discovery stage. In Figure \ref{fig:su4} an 
ATLAS low mass SUSY study is demonstrating the 
kinematic edge of the dilepton invariant mass $M_{\ell\ell}$. 
The edge is a measure of mass differences between the 
sparticles that are involved in
the decay (here the $\tilde{\chi^{0}_{2}}, \tilde{\ell_{R}} 
~{\rm and}~ \tilde{\chi^{0}_{1}} ~{\rm with }~$
$ M_{\ell\ell}^{max}=  M(\tilde{\chi^{0}_{2}}) \sqrt{1- \frac{M^{2}(\tilde{\ell_{R}})}
{M^{2}(\tilde{\chi^{0}_{2})} }}  \sqrt{1- \frac{M^{2}
(\tilde{\chi^{0}_{1}})}{M^{2}(\tilde{\ell_{R}})}})$. Similar edges 
are shown in Figure \ref{fig:clues} from CMS and ATLAS in 
different parts of the mSUGRA parameter space. 

Note that top, bottom, $Z$ and $W$ in the decays of sparticles (i.e. non-direct Standard Model production) in leptonic final states can also point towards rates and mass hierarchies of the SUSY (or other BSM) particles produced.

\section{The LHC SUSY Search, Orientation and Navigation Tool-Kit}
\subsection{Excesses as a function of luminosity}

The CMS and ATLAS  collaborations have published their physics performance reviews \cite{Ball:2007zza}, \cite{ATLAS-PTDR}. A rough summary of the 5$\sigma$ reach and the corresponding channels/analyses are given below (using the results from the most recent available results) in a format of what a publication might look like if/when such an excess is observed \footnote{There is a level of absurdity in the listing as presented here, however it is illustrative of the daunting task that the experiments will be faced with  when trying to interpret and cross-interpret the possible variety of signals they might observe, as these emerge; note that the luminosity values in parenthesis are rounded for the purposes of illustration and calculated with assumptions on the systematics - sometimes conservative and always referring to an understanding of the detector with 1 fb$^{-1}$.}:
\begin{small}
\begin{itemize}
\item  Search for SUSY (Evidence for excess) in $\ge$1 lepton + $E_{\rm T}^{\rm{miss}}$ + jets at 14 TeV  in the electron and muon channels (100 pb$^{-1}$).  

\item Search for SUSY (Evidence for excess)  in opposite sign dilepton pairs + $E_{\rm T}^{\rm{miss}}$ + jets at 14 TeV  in the electron and muon channels (20 pb$^{-1}$)
\item Search for SUSY (Evidence for excess) in same-sign dilepton pairs + $E_{\rm T}^{\rm{miss}}$ + jets at 14 TeV  in the electron and muon channels (200 pb$^{-1}$)
\item Search for SUSY  (Evidence for excess) in $Z^{0}$ leptonic decays+ $E_{\rm T}^{\rm{miss}}$ + jets at 14 TeV  in the electron and muon channels (100 pb$^{-1}$)

\item Search for LVF SUSY (Evidence for excess) in $e+\mu$ final state at 14 TeV (500  pb$^{-1}$)
\item Search for SUSY  (Evidence for excess)  in trileptons  at 14 TeV.  ($\sim \mathrm{fb}^{-1}$)

\item Search for SUSY (Evidence for excess) in   0 lepton + $E_{\mathrm T}^{\mathrm{miss}}$+ jets at 14 TeV  (10 pb$^{-1}$) 

\item Search for SUSY (Evidence for excess)  in $b\bar{b}$ + $E_{\rm T}^{\rm{miss}}$ + jets at 14 TeV (100 pb$^{-1}$) 

\item Search for SUSY (Evidence for excess)in top hadronic decays+ $E_{\rm T}^{\rm{miss}}$  at 14 TeV (200 pb$^{-1}$) 

\item Search for SUSY (Evidence for excess) in opposite-sign ditau +  $E_{\rm T}^{\rm{miss}}$ at 14 TeV 
(200 pb$^{-1}$) 

\item Search for GMSB (Evidence for excess) in prompt photon final states at 14 TeV (500 pb$^{-1}$) 

\item Search for GMSB (Evidence for excess) in non-pointing photons at 14 TeV  (1 fb $^{-1}$)

\item Search and reconstruction of heavy stable charged particles at 14 TeV
using TOF and dE/dx   (500 pb$^{-1}$) 
\item ....
\end{itemize}
\end{small}

Grouping the signatures for the sake of this discussion we still have a large listing of probable ``fast'' signals:

\begin{itemize}
\item {canonical inclusive} 
  \begin{itemize}
   \item  jets+ $E_T^{miss}$ (no lepton)
    \item jets+ $\ell$ + $E_T^{miss}$ 
    \item  same-sign dilepton + $E_T^{miss}$ 
    \item  opposite-sign same flavor dielectron and dimuon + $E_T^{miss}$ 
    \end{itemize}
 \item{higher reco object inclusive}
   \begin{itemize}
   \item  $Z$ + $E_T^{miss}$ 
   \item  $t$ hadronic + $E_T^{miss}$ 
   \item  $h^{0}(b\bar{b})$ + $E_T^{miss}$ 
   \end{itemize}
\end{itemize}

\begin{figure}[htb]
\begin{center}
    \includegraphics[width=.45\textwidth]{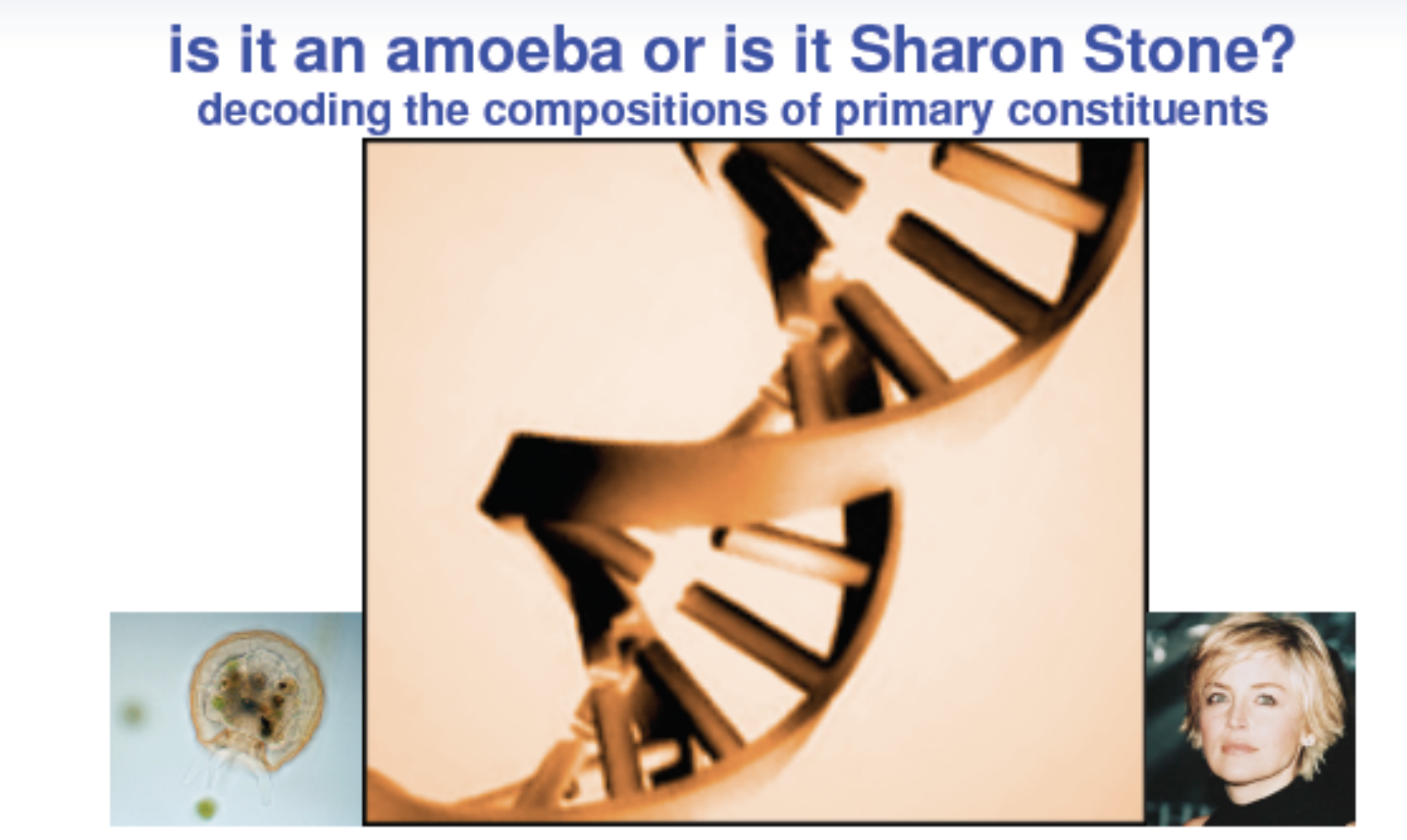}
\end{center}
    \caption{Just like decoding DNA we have to decode the signals we will observe. And we do expect more similarities than differences, so fast discrimination will require smart and simple measurements}
    \label{fig:amoeba}
\end{figure}

The matter in question is how exactly do we disentangle the emergent patterns in the observations (if/when excesses are observed) in order to get a direction towards the underlying mechanisms beyond the standard model. I like to depict this graphically in the form of Figure \ref{fig:amoeba}.

\begin{figure}[htb]
\begin{center}
    \includegraphics[width=.45\textwidth,height=.35\textwidth]{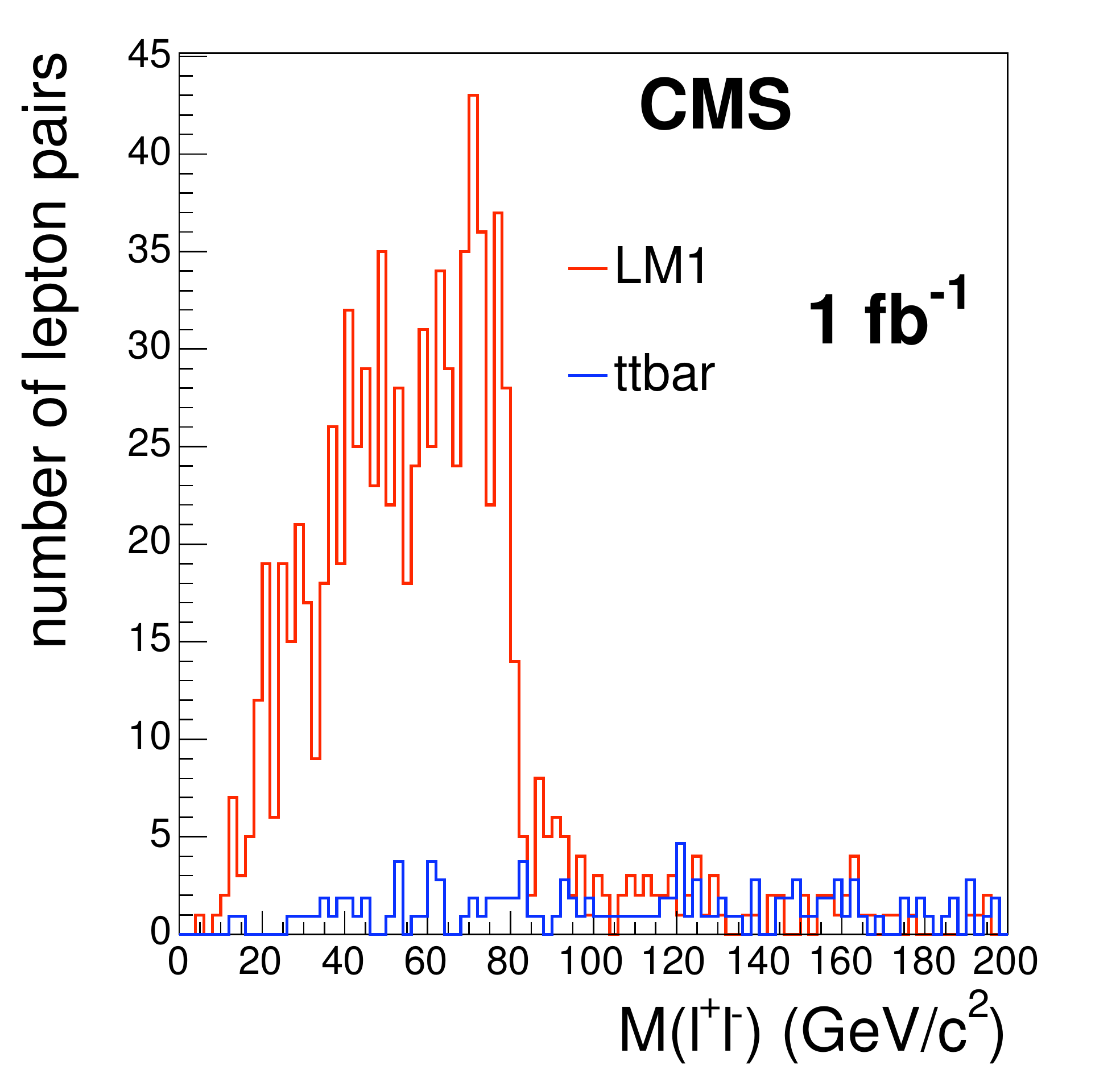}
    \includegraphics[width=.45\textwidth,height=.35\textwidth]{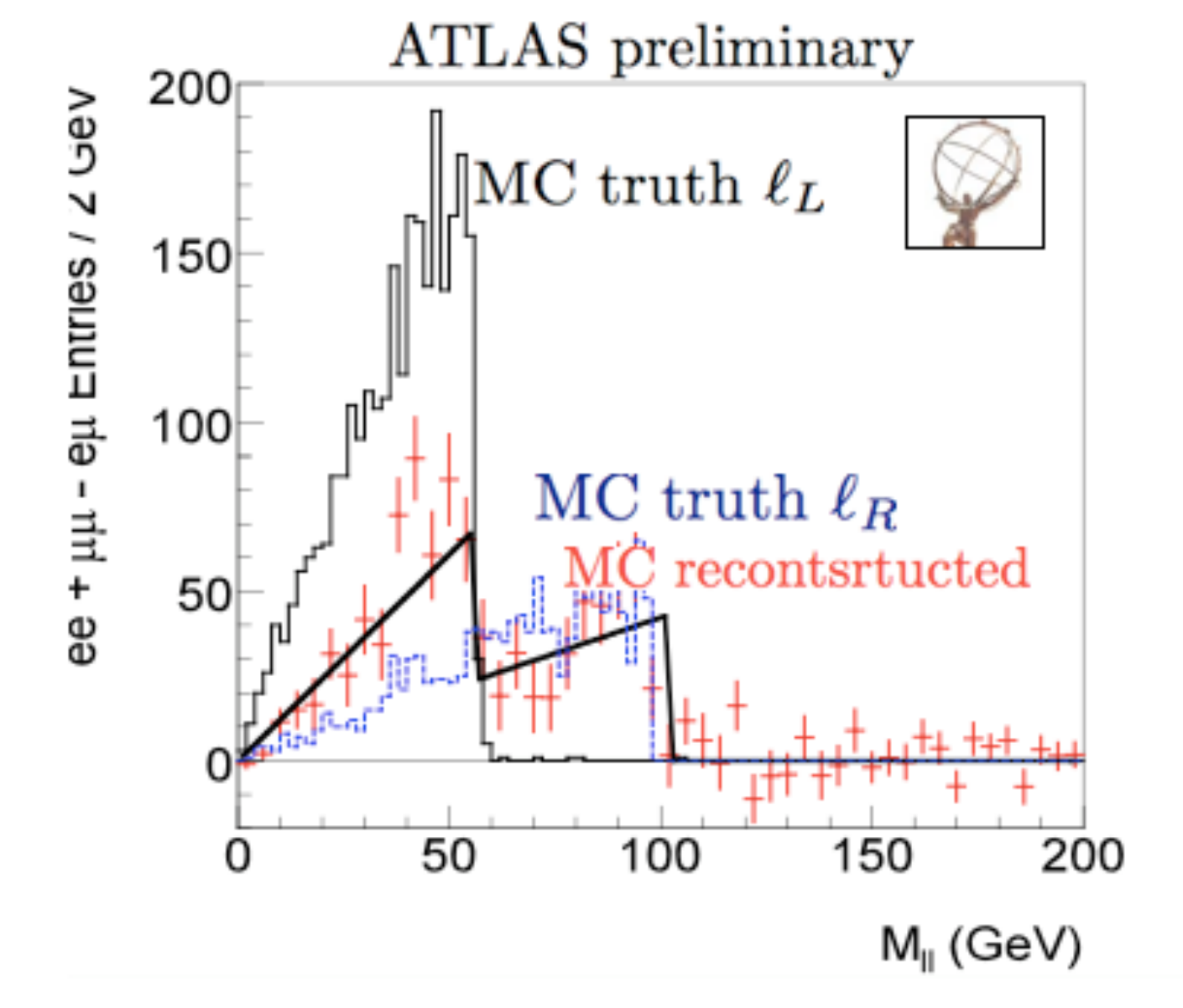}
    \includegraphics[width=.40\textwidth,height=.35\textwidth]{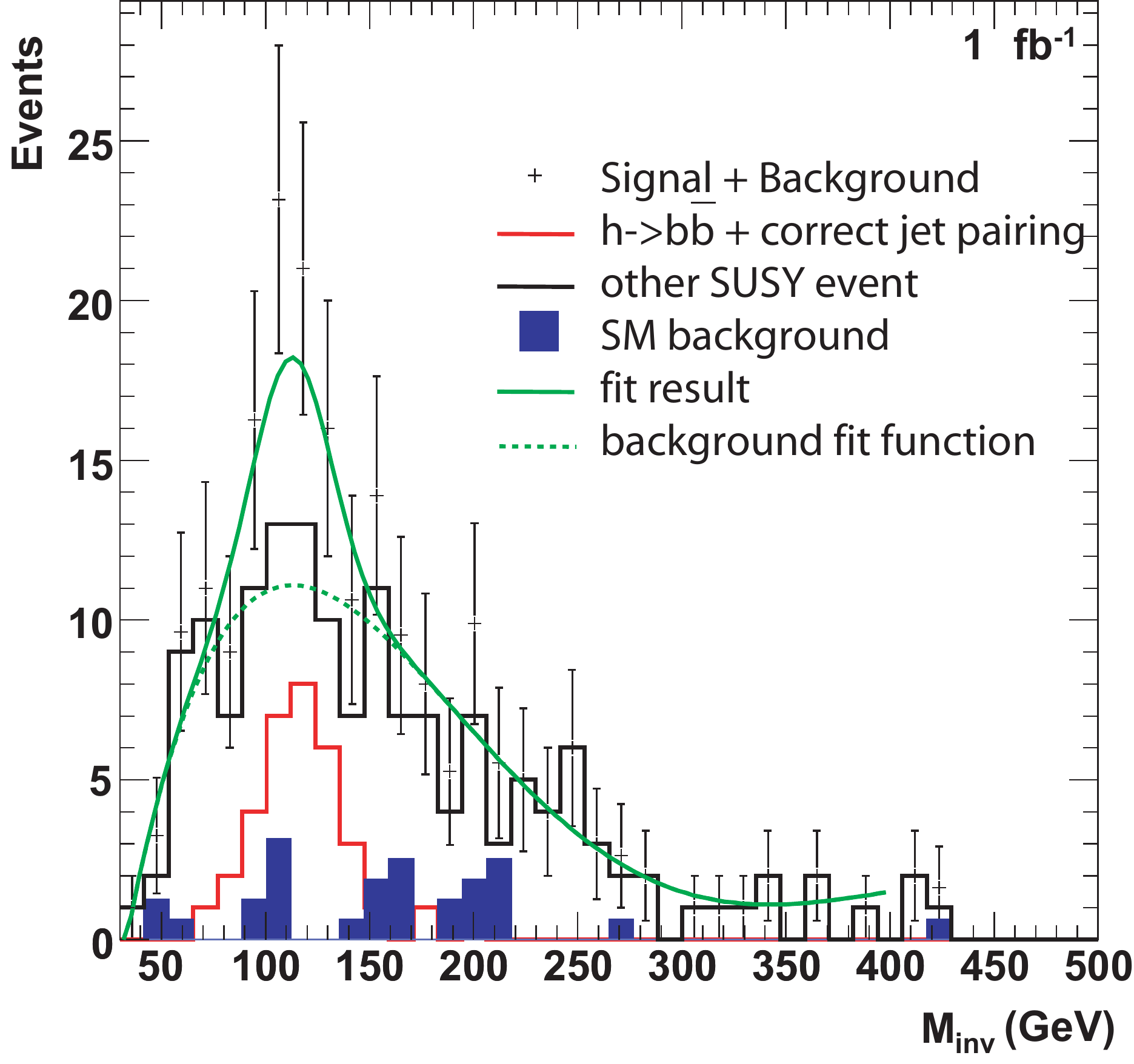}
\end{center}
    \caption{How will we use different observations to navigate the parameter space at start-up? What are the optimal measurables at start-up that will help piece-together a direction? (top) dilepton invariant mass from a CMS SUSY benchmark point analysis, (middle) similar from ATLAS, (bottom) $b\bar{b}$ invariant mass from a CMS SUSY benchmark point analysis.}
    \label{fig:clues}
\end{figure}

The question  is synonymous to the ``inverse LHC problem'' attacked with``footprint'' approaches \cite{kane}, MARMOSETs \cite{marmoset} and other strategies that include full-event harvest \cite{Gunion}, multivariate sophisticates analyses with decision trees\cite{DT}, spin-prints \cite{Barr0}, \cite{Barr1}) as well as systematic understanding of the SUSY available kinematics and topologies  \cite{luc},  and defining a strategy for distinguishing ``look-alike'' variations within SUSY itself and other frameworks \cite{ll}. I give in Figure \ref{fig:clues} a set of possible reconstructed mass edges and ``bumps'' that might emerge with early data at ATLAS and CMS.

\begin{figure}[htb]
\begin{center}
    \includegraphics[width=.40\textwidth,height=0.70\textwidth]{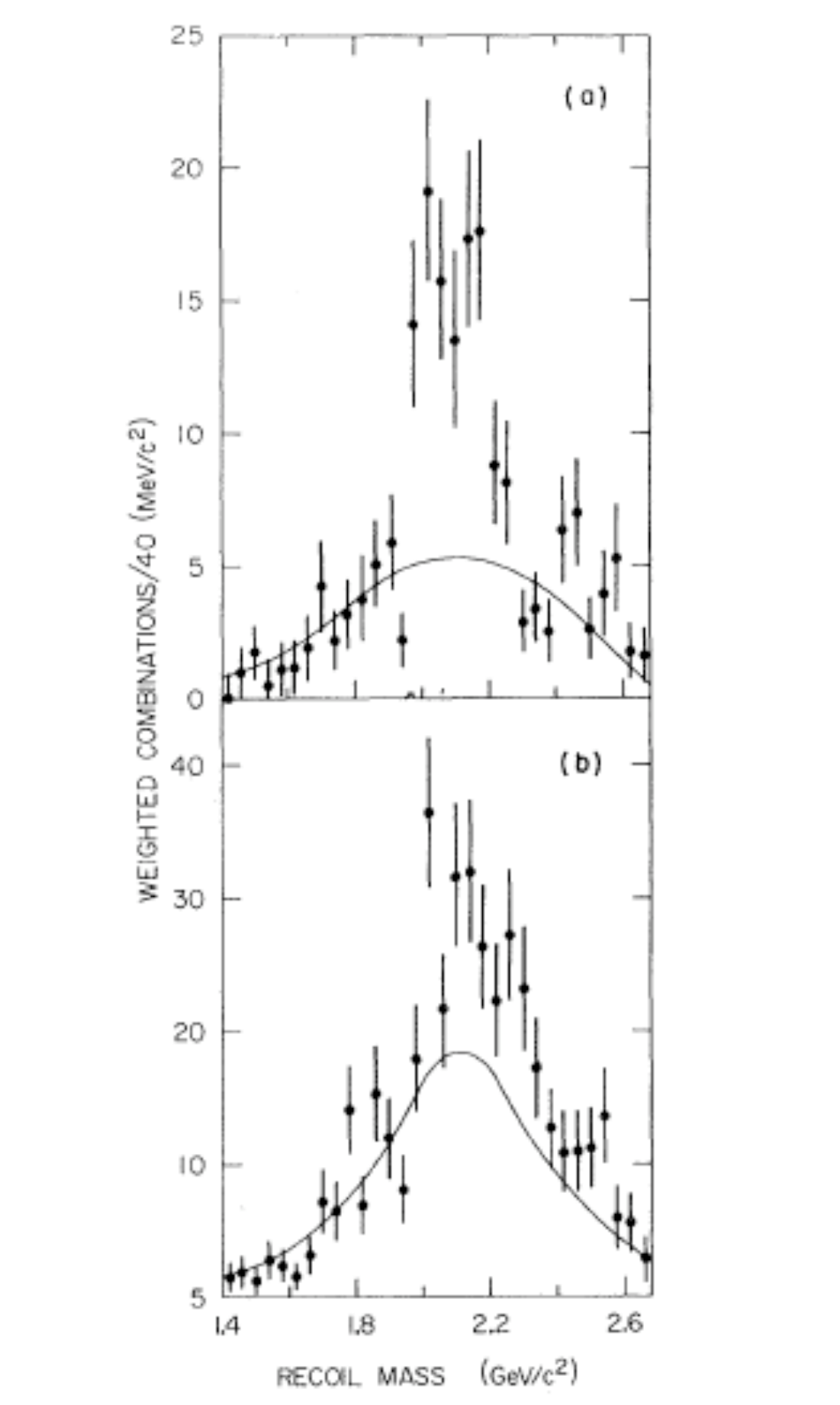}
\end{center}
    \caption{Recoil spectra for combination in the $K\pi$ and $K\pi\pi\pi$ peaks from \cite{Goldhaber:1976xn}. }

    \label{fig:charm}
\end{figure}

In figure \ref{fig:charm} I give the recoil mass spectrum associated with the then (1976) newly discovered charmed mesons in $e^{+}e^{-}$ annihilation at SPEAR  \cite{Goldhaber:1976xn}.  The study, interpretation and predictions based on
these measurements were published concurrently \cite{DeRujula:1976xn} and involved threshold, form factor, spin-effects and mass splitting analysis. The interpretation template then was the charm hypothesis.
As Michelangelo Mangano pointed out this is possibly the closest Standard Model example  that could illustrate the least we foresee to be faced with regarding 
discoveries and patterns to be interpreted with the early data 
at the LHC.  One of the difficulties now is that the interpretation templates are infinite.

Nevertheless, the study of the questions of the type that follows could point us to a direction:

\begin{itemize}
\item if excess of SS dileptons $\rightarrow$  ?
\item if $++/--$ =2 $\rightarrow$  ?
\item if excess if OS dileptons  $\rightarrow$  ?
\item if triangle in dilepton invariant mass  $\rightarrow$  ?
\item if double triangle $\rightarrow$ ?
\item if no triangle  $\rightarrow$ ?
\item if $Z^{0}$ and no triangle  $\rightarrow$ ?
\item if $Z^{0}$ and triangle  $\rightarrow$ ?
\item jet and lepton object counting and ratios (i.e. 3j/4j/5j/6j, 1$\ell$/2$\ell$/3$\ell$/4$\ell$),  $\rightarrow$ ?
\item ...  
\end{itemize}

\section{Conluding Remarks}

Our  current and extrapolated status-of-being 
as a field is summarized very eloquently in
this meeting by the introductory talk ``Anticipating a New Golden Age''
of Frank  Wilczek \cite{Wilczek:2007gsa}. 
I would like to make a few very obvious comments here.
\begin{enumerate}
\item Although we cannot predict the experimental data at the LHC we do
build a strong preparatory program of analysis strategies for the potential discovery physics search, navigation and orientation.
\item Within this preparatory program we observe anew the strong concilience between theory and experiment.
\item The emergent confluence between cosmological-- especially on the dark matter, and particle physics data presents us with a real reciprocity. The relic density for a given dark matter candidate cannot be directly measured, it must be calculated and this requires knowledge of its mass and its interactions that are relevant to how it annihilates in the early universe. Both the cosmology standard model and all the beyond the particle physics standard model scenarios
have large ``terra incognita'' sectors: the exercise of constraining cosmology
using assumed beyond the standard model physics frameworks (and data from direct DM searches) and vice-versa will be a major part of the physics program at the LHC (see also the DM discussion in \cite{Wilczek:2007gsa}). 
\item Based on our current knowledge, supersymmetry is the most plausible theory to extend the Standard Model in the TeV scale and should have  already been observed in the LEP, Tevatron or low energy data. The searches at the LHC use various subsets of supersymmetric points in the vast parameter space as templates to provide a signature space that is well studied in preparation for the much anticipated confusing multitude of SUSY-mutant features in the data (as opposed to studying a few points in the mSUGRA \footnote{Note that using the formulation of \cite{Hall:1983iz} mSUGRA is equivalent to the constrained CMSSM model, see also \cite{lykkenLP07}.} parameter space). 
\item Finally I would like to close with a few words of caution: In my talk at this meeting I showed how the different modern SUSY spectra calculators that we use in the LHC experiments give different results in particular corners of the parameter space and for  small variations of standard model input values (I used the top mass as an example). While there is a lot of progress, there is also a lot of work remaining for a consistent implementation of SUSY mutli-body decays and SUSY QCD associated production. Similar caution was raised by Michelangelo Mangano \cite{MLM} on most all the standard model QCD associated production. We will use the data to calibrate the standard model but it is important to design exactly how we will do this given that the data will be contaminated by discoveries.
\end{enumerate}

\section*{Acknowledgements}
I would like to thank especially Wim De Boer, Luc Pape, Shoji Asai, David Costanzo, Michelangelo Mangano, Gordy Kane, Joe Lykken and the CMS and ATLAS collaborations.

%
\bibliographystyle{h-physrev.bst}
\bibliography{smaria}
 
\end{document}